\pdfoutput=1
\documentclass[numberedappendix]{emulateapj}
\usepackage[breaklinks,colorlinks,citecolor=blue,linkcolor=magenta]{hyperref}
\usepackage{psfig}
\usepackage{amsmath}

\begin{document}

\title{Spectral Hardening in Black Hole Accretion: Giving Spectral Modelers an \MakeLowercase{f}.}

\author
{Shane W. Davis\altaffilmark{1},
and Samer El-Abd\altaffilmark{1}}

\affil{$^{1}$Department of Astronomy, University of Virginia, Charlottesville, VA 22904, USA}

\email{swd8g@virginia.edu}

\begin{abstract}

By fitting synthetic spectral models computed via the TLUSTY code, we
examine how the spectra from thin accretion disks are expected to vary
in accreting black hole systems.  We fit color-corrected blackbody
models to our synthetic spectra to estimate the spectral hardening
factor $f$, which parameterizes the departure from blackbody and
is commonly used to help interpret multitemperature blackbody
fitting results.  We find we can define a reasonably
robust $f$ value to spectra when the effects of Compton scattering
dominate radiation transfer.  We examine the evolution of $f$
with black hole mass and accretion rate, typically finding a moderate variation
($f \sim 1.4-2$) for accretion rates between 1\% and 100\% of the Eddington rate.
Consistent with most previous work, we find $f$ tends to increase with accretion rate, but
we also infer a weaker correlation of $f$ with black holes mass.  We find
that $f$ is rarely much larger than 2 unless the disk becomes photon
starved, in contention with some previous calculations.  Significant
spectral hardening $(f > 2)$ is only found when the disk mass surface
density is lower than expected for $\alpha$-disk models unless
$\alpha$ is near unity or larger.
\end{abstract}


\section{Introduction}

It is widely believed that the soft X-ray spectral component in black
hole X-ray binaries is thermal emission from an optically thick,
geometrically thin accretion disk \citep{ShakuraSunyaev1973}.  This
component is generally present in high count rate phases of X-ray
transients in low mass X-ray binaries, when the source is said to be
in a high/soft (thermal dominant) or steep power-law (very high,
intermediate, soft/intermediate) state \citep{RemillardMcClintock2006,Doneetal2007}.  Spectral modeling
of this component can (in principle) tell us many things about the
properties of the accretion flow and black hole.  We can look for
changes in the flow structure and geometry as other properties
(e.g. the luminosity) of the source varies.  If the spectrum of this
component and its variations are well-matched by an accretion disk
model based on numerical simulations or analytic calculations, then it
may even be possible to infer properties (e.g. spin) of the black hole
itself \citep{McClintocketal2011}.

The simplest way to turn an accretion flow model into a spectrum is to
assume blackbody emission from the photosphere of an optically thick
disk.  A range of disk models with varying levels of complexity and
different disk structures have adopted this assumption, with the
DISKBB model \citep{mit84} from Xspec \citep{Arnaud1996} being the most widely used to model accretion
disk emission.  However, it is well understood that electron scattering
plays an important role in radiation transfer at the characteristic
photon energies and temperatures, and generically leads to deviations
from blackbody emission.  Hence, one might expect blackbodies to be a
generically poor approximation in this
limit.  Fortunately, the exchange of energy between electrons and
photons that are inelastically scattered (i.e. Compton scattered)
generally enforces a Wien tail at the high energy end of the
spectrum \citep[][ST95 hereafter]{ShimuraTakahara1995}.  ST95 showed
that the resulting spectra can be approximately modeled by a color-corrected (or diluted) blackbody
\begin{equation}
I_\nu=\frac{2 h}{c^2 f^4} \frac{\nu^3}{\exp\left(\frac{h \nu}{f k T}\right)-1},\label{eq:ccbb}
\end{equation}
where $I_\nu$ is the specific intensity, $\nu$ is frequency, $k$ is
Boltzmann's constant, $h$ is Planck's constant, $c$ is the speed of
light, and $T$ is the temperature.  The key difference from a normal
blackbody is presence of $f$, which is commonly referred to as the
color correction or spectral hardening factor.  (For the remainder of
the paper, we will refer to this spectral shape as a color-corrected
blackbody but refer to $f$ as the spectral hardening factor.)  It is
the multiplicative factor by which spectral features are shifted to
higher energies.  The factor of $f^{-4}$ keeps the frequency integrated flux
fixed.

This concept is particularly useful when paired with a multicolor (or
multitemperature) disk blackbody model with a power law dependence of flux on
radius.  The most popular example is the DISKBB model in Xspec, which
assumes that flux $F$ scales with radius $r$ according to $F \propto
r^{-3}$.  This self-similar model only has two free parameters, a
normalization and the innermost disk temperature.  If one assumes all
emission regions have the same associated $f$ value, then one can use
DISKBB to fit for the innermost disk temperature and treat the best
fit value as a type of color temperature, which is divided by $f$ to
obtain an effective temperature.  Hence, a common use of $f$ is to
color-correct the best-fit temperature, which can be used along with
the best fit normalization to obtain an approximate measurement of the
inner disk radius.  This radius is sometimes compared with the innermost
stable orbit of a spinning black hole to estimate black hole spin \citep[see e.g.][]{Zhangetal1997}.

Of course, all this relies on having some mechanism for computing $f$.
Efforts have been made to ``empirically''\footnote{We put empirically
in quotes because these estimates makes implicit assumptions about the
emission/scattering geometry through their choice of Comptonizing model. Although we do
not view the models as empirical estimates for this reason, they still provide a useful,
indpendent constraint on $f$.} measure
$f$ \citep{Cuietal2002}, by fitting for the ratio of seed
photon energies to the final color temperature, which is obtained by
fitting a particular choice of model (e.g. COMPTT in Xspec).
For example, \cite{PszotaCui2007} infer $f \sim 1.3-1.5$,
broadly consistent with models discussed below.

Several attempts to theoretically compute $f$ have been made,
beginning with ST95, who compute one dimensional models of the
vertical structure of the accretion disk atmosphere while solving
radiation transfer with the effects of Compton scattering included.
ST95 found that $f$ varies only over a relatively small range (1.7-2)
for X-ray binaries accreting at greater than a few percent of the
Eddington rate. These results were broadly supported by the work
of \citet[][hereafter D05]{Davisetal2005}, who utilized the TLUSTY
stellar atmospheres code \citep{Hubeny1990,HubenyLanz1995} to compute
models of the vertical structure and radiation transfer in the
accretion disk annuli.  D05 found a consistent values for
spectral hardening factors over the range considered by ST95, with a
slight increase in $f$ as accretion rate increased.  In contrast, the
models of \citet[hereafter MFR00]{Merlonietal2000} found a substantial
increase in $f$ in models where a fraction of the dissipated energy
was assumed to take place in a corona, leaving only a small amount of
flux in the disk.  Although D05 did not consider the impact of
dissipation in a corona, a comparison of the spectral hardening
factors at the same effective accretion rate find much lower spectral
hardening factors for the D05 models.

Although $f \sim 1.7$ is still commonly utilized for X-ray binaries,
there are several motivations to better understand the range of $f$
found in accretion disk models.  As mentioned above spectral hardening
factors can be used in relativistic models of black hole accretion
disk to provide spin estimates based on continuum
fitting \citep{McClintocketal2011}.  Large tables of spectral
hardening factors have been computed for this
purpose \citep{McClintocketal2006}, but their general properties have
not been reported on previously.

A second consideration comes from
observations of X-ray binaries in low hard states.  Although the
longstanding picture is that the inner edge of the disk moves out in
these states \citep{Esinetal1997}, this notion has been challenged by
spectral fits suggesting relativistically broadened Fe K$\alpha$
emission and inner disk temperatures consistent with the disk extending
nearly to the innermost stable circular
orbit\citep{ReynoldsMiller2013}.  Such observations might be
reconciled if the spectral hardening factor could increase
significantly in these states, in a manner similar to the results of
MFR00.

A third consideration is the potential evolution of $f$ for
accretion disks at larger masses, such as in the intermediate mass
black hole regime \citep{hkh05,Davisetal2011} or in active galactic nuclei (AGN)
regime.  Although the
color-corrected blackbody is not generally expected to provide a good
fit for the optical to UV emitting regions of accretion disks, where
Compton effects are thought to be small \citep[e.g.][]{hub01}, they
may provide a good approximation to the soft X-ray emission coming
from the hottest, innermost regions of relatively low mass and high
accretion rate AGNs.

Hence the goal of this paper is to explore the range of $f$ inferred
over a wider range in disk parameters encompassing models with a
larger range of accretion rates and masses.  As a basis for this study,
we utilize TLUSTY based models of accretion disk annuli that
underlie the BHSPEC model \citep{DavisHubeny2006}, but have also been used with the KERRBB2 model \citep{McClintocketal2006} and slim disk models \citep{Straubetal2011}.

\section{Model and Methods}

The spectral models of Comptonized accretion disks employed in this
study were computed using the TLUSTY stellar atmospheres
code \citep{Hubeny1990,HubenyLanz1995}.  The properties of these
models are discussed in D05 and \citet{DavisHubeny2006} and we refer
the reader to these earlier works for more information.

The TLUSTY disk models each represent an individual annulus in an
axisymmetric accretion disk.  The models are computed assuming
physical quantities (density, temperature, radiation flux, etc.) only
vary as a function of height within the annulus.  The lower boundary
of the model corresponds to the disk midplane and the upper boundary
corresponds to a region of low optical depth above the photosphere.  Models
are computed by simultaneously solving the equations of hydrostatic
equilibrium, statistical equilibrium for level populations,
conservation of energy, mass conservation, and radiation transfer.
These are solved using iterative methods (complete linearization and
accelerated lambda iteration) commonly employed in stellar atmosphere
calculations \citep{HubenyLanz1995}.

\begin{figure}[h]
\begin{center}
\includegraphics[width=7.6cm]{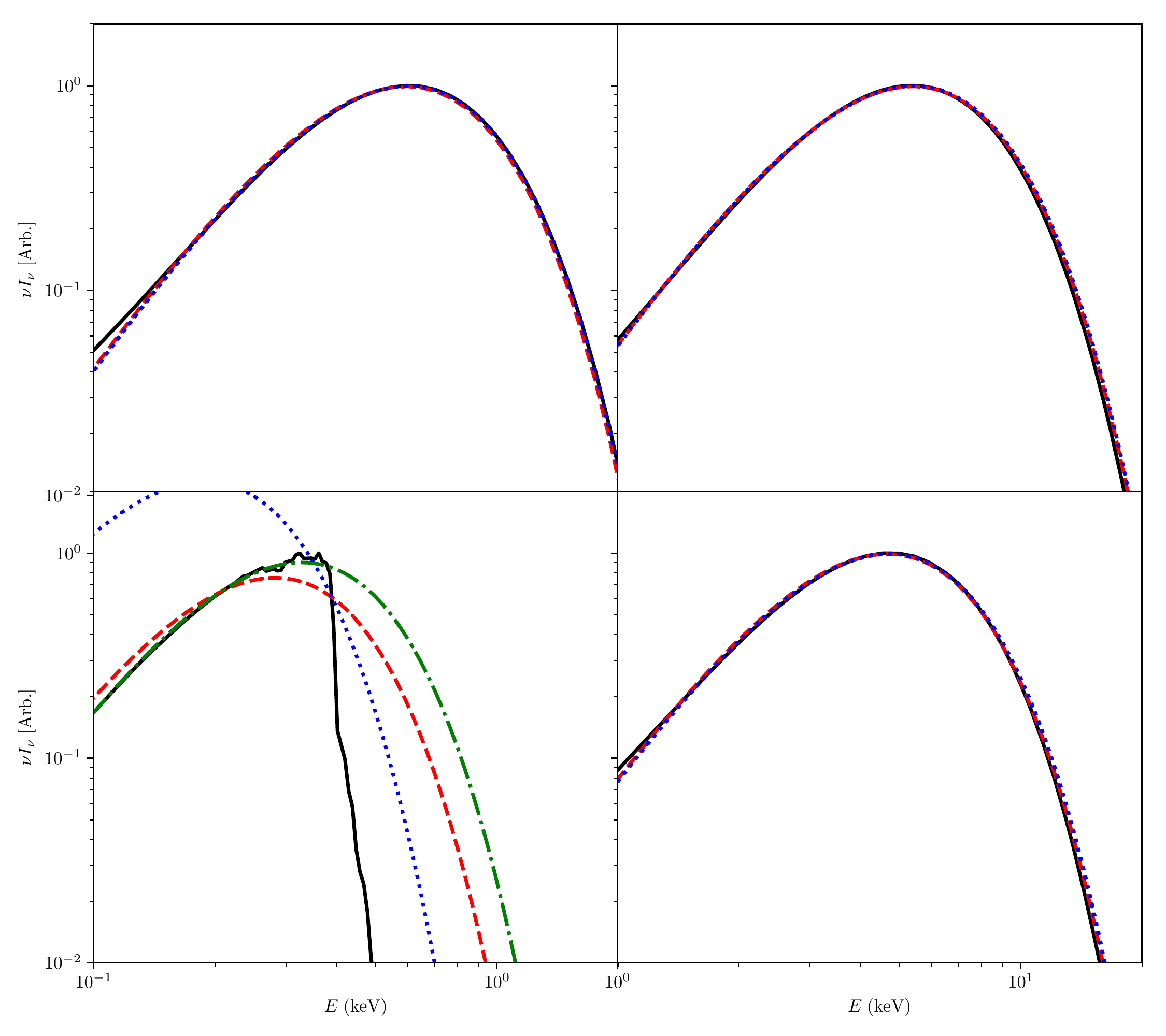}
\end{center}
\caption{Specific intensity versus photon energy for four different spectral models corresponding
to $(\log T_{\rm eff}, \log m_0, \log Q)$ of (5.8, 3, -3) (upper
 left), (6.9, 3, 6) (upper right), (5.8, 3, 6) (lower left), and (6.9,
 5, 6) (lower right).  Each panel is plotted for an inclination near
 $60^\circ$ and also showed the best-fit colored corrected blackbody
 fit with the unweighted (red, dashed), weighted, absorbed (blue,
 dotted), and weighted, unabsorbed (green, dot-dashed) methods.
 Intensities are plotted with an arbitrary renormalization.  Note that
 the range of the horizontal axis differs for the left and right
 panels.}
\label{f:spec4}
\end{figure}

Hydrostatic equilibrium is computed using an equation of the form
\begin{equation}
-\frac{d P_{\rm tot}}{d z} = \rho Q z,
\end{equation}
where $z$ is the height above the midplane ($z=0$ at the midplane),
$P_{\rm tot}$ is the sum radiation and gas pressure, $\rho$ is the
mass density, and $Q$ is a parameter that characterizes the strength
of vertical (tidal) gravity.  For a Newtonian disk
$Q \simeq \Omega^2$, where $\Omega$ is the Keplerian angular rotation
rate within the disk, but can differ when general relativistic
effects are included.  It is often convenient to replace derivatives
with respect to $z$ by the column mass $m$, defined via $d m = - \rho
dz$.  The TLUSTY models are computed on a grid in $m$ ranging from $m
= 10^{-3} \; \rm g/cm^2$ to $m_0$, the maximum column mass at the
midplane.

Since each model represents an individual annulus, they are
determined by local (i.e. radially varying) disk parameters: the
mass surface density $\Sigma$ ($\Sigma = 2 m_0$, so we will use $m_0$
instead of $\Sigma$ hereafter), the radiative flux $F$ or effective temperature $T_{\rm
eff}=(F/\sigma_{\rm sb})^{1/4}$ at the surface, and the tidal gravity parameter $Q$.  A global
accretion disk model (such as the $\alpha$-disk) can be used to
specify $m_0$, $T_{\rm eff}$, and $Q$ as functions of radius $r$ and
global parameters black hole mass $M$, black hole spin $a$, and
accretion rate $\dot{M}$.  Our TLUSTY models are tabulated in $m_0 \,
\rm (g/cm^2)$, $T_{\rm eff} \, \rm (K)$,
and $Q \, \rm(s^{-2})$. We compute $\log Q$ ranging from -10 to 11, in
steps of 1 dex, $\log T_{\rm eff}$ from 5 to 7.5 in steps of 0.1 dex,
and $\log m_0$ at values of 2.5, 2.75, 3, 4, 5, and 6.  This range us
allows us to characterize the X-ray emitting regions of black hole
accretion flows for black hole masses ranging from $\sim 3 M_\odot$ to
$10^9 M_\odot$ Each model spectrum is computed and stored as the value
of the specific intensity $I_\nu$ on a grid evenly space in $\log \nu$
and $\cos \theta$, where $\nu$ is the frequency and $\theta$ is the
inclination measured relative to the surface normal.  Spectra are
tabulated on a grid containing 350 frequencies and 10 angles.

\section{Results}

\subsection{Determining the Spectral Hardening Factor}
\label{fitting}

Since the color-corrected blackbody is only an approximation
(sometimes a poor approximation), there is no unique method for
determining $f$ for the TLUSTY models.  One way to associate an $f$
with the models is to fit the color-corrected blackbody form to the
model specific intensity as a function of photon energy, following
ST95. However, this fit will necessarily depend on assumptions about
the energy range fit and the relative weighting or uncertainty used
for different photon energies.  For example, one could assume each
photon energy in the model has the same error or assume that the error
follows photon statistics for a photon count rate within well defined
energy bins.

Previous efforts to estimate $f$ have employed realistic instrumental
response matrices and accounted for the impact of interstellar
absorption to closely approximate the circumstances of real
observations and were obtained by fitting global disk models
\citep{McClintocketal2006}.  Since we are generally interested in the hardening factor
associated with local annuli models, we prefer to avoid adopting any
instrument specific response and instead use simpler prescriptions.
Although the precise value of $f$ will be sensitive to any specific
prescription, we are predominantly interested in the variation in $f$
with physical parameters, which we expect to be less sensitive to the
details of our fitting procedure.  In order to test this, we consider
three different fitting models and use linear least squares fitting in
each case.  We denote TLUSTY intensities as $I_i$ on a grid of
frequencies $\nu_i$, where $i$ runs over the frequency grid and the
function $I_{\rm cc}(f,T,\nu)$ is the function given by equation
(\ref{eq:ccbb}).

\begin{figure}[h]
\begin{center}
\includegraphics[width=7.6cm]{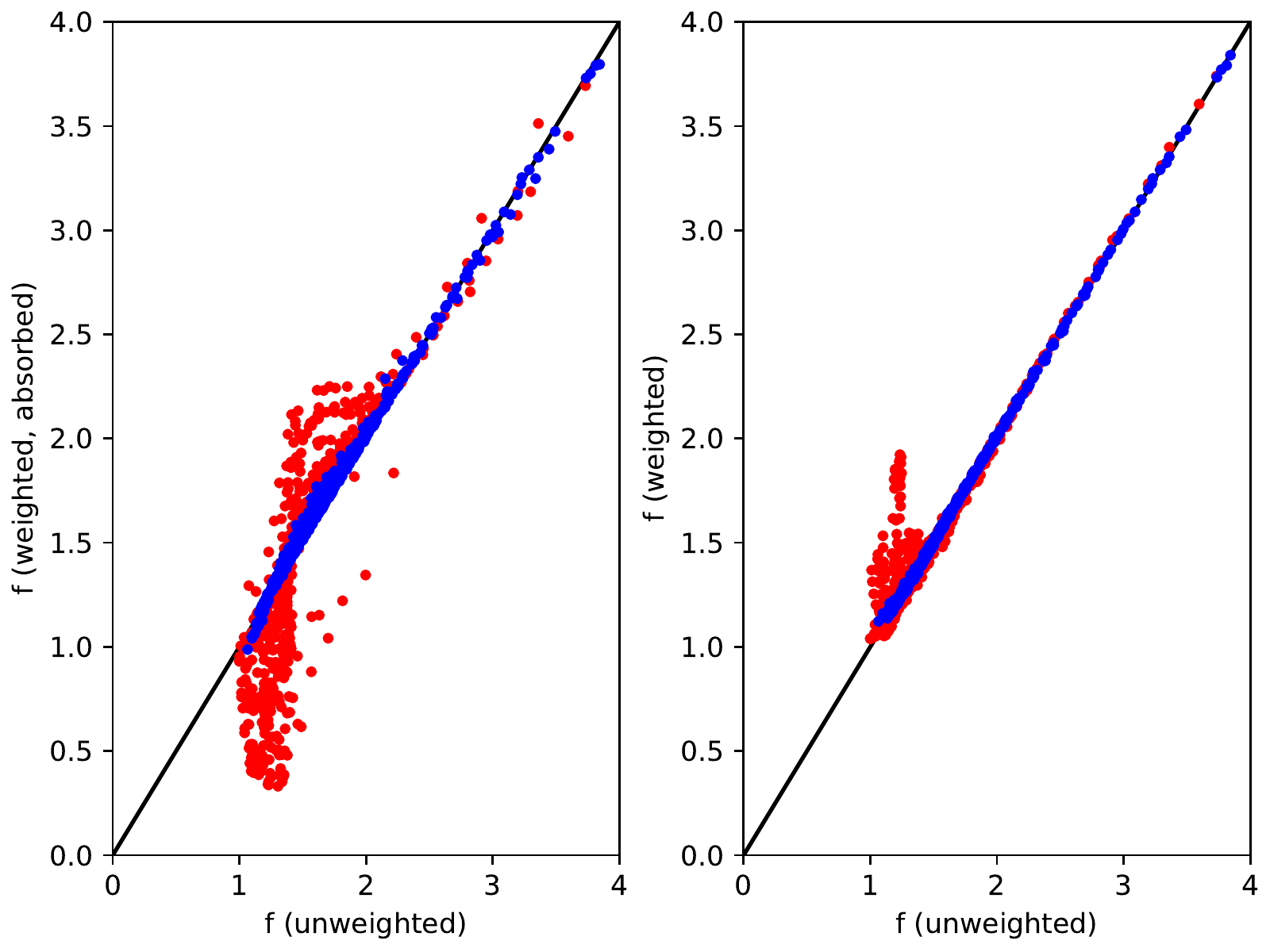}
\end{center}
\caption{Comparison of best-fit spectral hardening factors $f$ for different
fitting methods.  In both panels, the horizontal axis correspond to the value
of $f$ obtained from fitting using the unweighted method.  The vertical
axis on the left and right panels are the best-fit $f$ obtained with
the weighted absorption and weighted (no absorption) methods, respectively.
Each point correspond to a different annulus for a viewing angle of $60^\circ$.
Points are color-coded by where the spectrum peaks in $I_\nu$, with blue
circles representing models with spectral peaks above 0.5 keV and red
symbols representing those peaking below 0.5 keV.
and p}\label{f:fcomp}
\end{figure}

The first method, which we refer to as the unweighted method, is to
simply assume each intensity has the same uncertainty and minimize the
difference $I_i-I_{\rm cc}(f,T_{\rm eff},\nu_i)N$, where
$T_{\rm eff}$ is the effective temperature of the annulus being fit.
$N$ is a normalization parameter that accounts for the fact that the
spectra can be rather anisotropic (limb darkened) due to the effects
of electron scattering.  This gives us two best-fit parameters, $N$
and $f$, for each angle (fit separately) for each model annulus.
Typical values of $N$ range from 0.8 to 1.2 and are generally
consistent with expectations from limb darkening, with $N$ being
larger than one for face-on inclinations and lower for edge on inclinations.
We will focus on $f$ in this work.  The best fit models using this
method are shown as red-dashed curves in the panels of
Figure~\ref{f:spec4}.  In all but the lower left panel, the
color-corrected blackbody provides a reasonably good approximation to the
model spectrum near the peak.  In the lower left panel, which
represents an annulus with a relatively cool photospheric temperature,
absorption edges are prominent and the spectrum deviates strongly from
the color-corrected blackbody approximation.  This unweighted method
tends to underestimate the flux near the peak because frequencies near
the spectral peak receive the same weight as frequencies in the
exponential tail.

\begin{figure*}[t]
\begin{center}
\includegraphics[width=15cm]{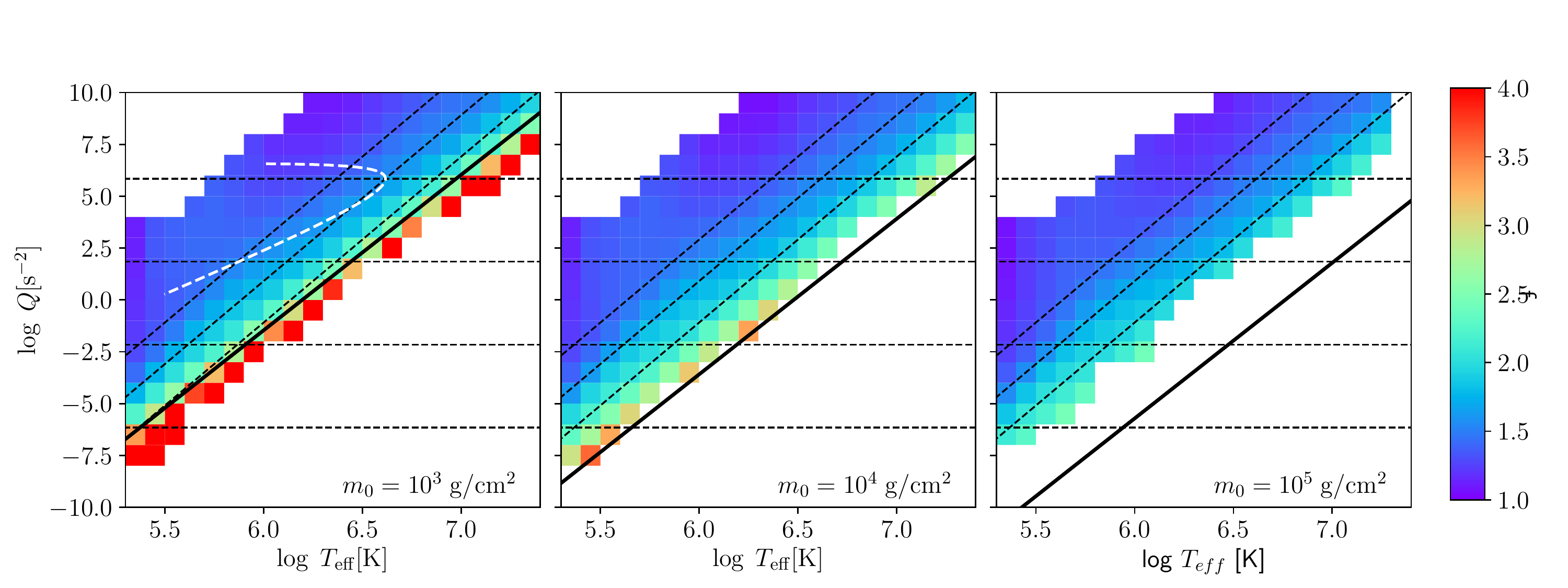}
\end{center}
\caption{Table of best-fit spectral hardening factors $f$.  The vertical axis shows the log of tidal parameter $Q$($=\Omega^2$ for Newtonian) and the horizontal axis show the log of the effective temperature $T_{\rm eff}$. Each square represents one model. From left-to-right the panels are evaluated for $m_0=10^3$, $10^4$, and $10^5$ $\rm g/cm^2$, respectively. The dashed diagonal lines represent curves of constant Eddington ratio ($\dot{m}=0.01, 0.1, 1$ from left to right) and horizontal dashed curves show lines of constant mass ($M/M_\odot=10, 10^3, 10^5, 10^7$ from top to bottom) evaluated for the hottest annulus in the disk model.  The white dashed curve in the left panel shows the $Q-T_{\rm eff}$ variation as a function of radius for a relativistic $\alpha$-disk model with $a=0$, $M=10 M_\odot$, and $\dot{m}=0.1$. The thick black solid curve corresponds to an analytical estimate (equation [\ref{eq:starved}]) of where the annulus becomes photon starved.  Note that we cap the color bar at $f=4$ to better show the variation at lower values of $f$, but several of hardest spectral models exceed this value.}\label{f:table3}
\end{figure*}

In more realistic situations, photon statistics and estimates of
systematic error in the detector determine the uncertainty in each
ordinate $y_i$.  Here we simply approximate the relative error due to
photon statistics as $\sigma_i=1/\sqrt{y_i}$.  This is approximately
what one would find by integrating photon number $(I_\nu/(h\nu)$ over
bins that logarithmically spaced in photon energy.  Our second method,
which we refer to as the unabsorbed weighted method, simply assumes
$y_i = I_i$ and again minimizes $I_i-I_{\rm cc}(f,T_{\rm
eff},\nu_i)N$.  A downside of this method is that it does not account
for the fact that typical observations are performed in the X-rays
where interstellar absorption can significantly attenuate the signal
as soft X-ray energies below $\sim 2$keV.  To gauge the impact of
interstellar absorption, we introduce an attenuation factor $A_{\rm
abs}(\nu)$, which corresponds to the interstellar absorption predicted
by the Xspec PHABS model \citep{Arnaud1996} with $N_{\rm H}=10^{20} \; \rm cm^{-2}$.  We then
fit for the absorbed weighted model by minimizing $A_{\rm abs}(\nu_i)
(I_i-I_{\rm cc}(f,T_{\rm eff},\nu_i)N)$ with $\sigma_i=1/\sqrt{A_{\rm
abs}(\nu_i)y_i}$.  Hence, the same attenuation factor is applied
to both the model spectrum and its best fit color-corrected
blackbody.  In effect, this acts to strongly reduce the weight
of lower frequencies in the spectral fitting due to the quasi
exponential increase in attenuation as photon energy decreases.

\begin{figure*}[t]
\begin{center}
\includegraphics[width=15cm]{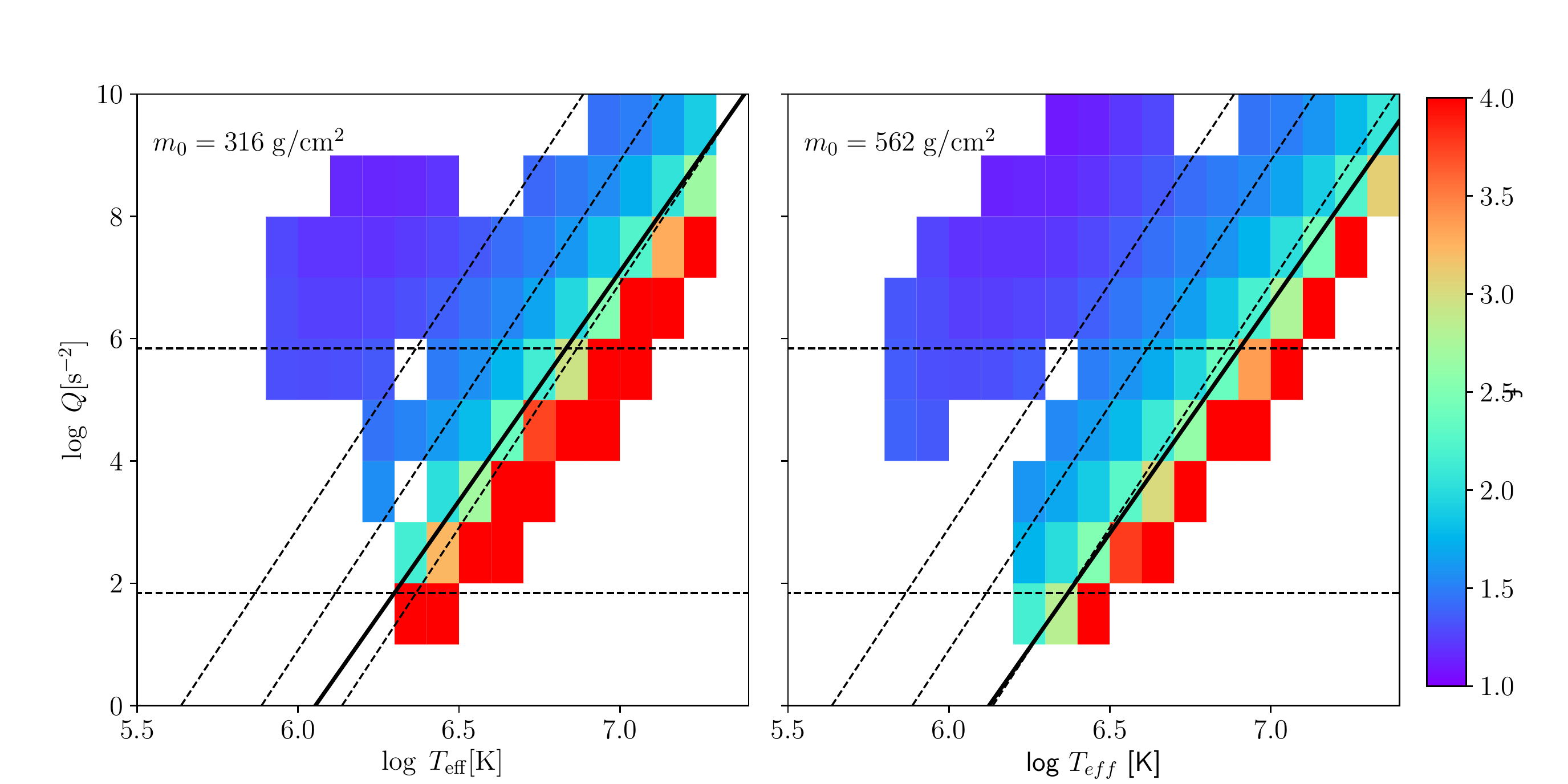}
\end{center}
\caption{Table of best-fit spectral hardening factors $f$.  All curves have the same meaning as in Figure~\ref{f:table3}, but are plotted for $m_0=10^{2.5} \; \rm g/cm^2$ (left) and  $m_0=10^{2.75} \; \rm g/cm^2$ (right). We cap the color bar at 4, although $f$ exceeds this value for some models.}\label{f:table2}
\end{figure*}

The weighted absorbed and weighted unabsorbed models are shown as
blue-dotted and green, dot-dashed curves (respectively) in
Figure~\ref{f:spec4}.  The best-fit spectra show a high level of
agreement in all but the lower left panel.  For the models where
the color-corrected blackbody provides a good fit, the best-fit
spectral hardening factors are in reasonably good agreement between
the different fitting methods.  However, when the color-corrected
blackbody is a poor fit, the best-fit $f$ depends sensitively on the
fitting method.  In the lower left panel, there is a prominent edge
due to N photoionization at 0.4 keV (as well as C and O features).
The unweighted method falls below the model
redward of the edge and exceeds the model blueward.  For the weighted
unabsorbed model, the smaller uncertainty near the peak forces a better
match just redward of the edge at the expense of greatly overpredicting
the intensity blueward of the edge.  The weighted absorbed model
provides a much lower $f$ with a huge excess at lower photon energies.
The poor fit at these energies has a low contribution to the overall fit
due to the effect of strong absorption leading to relatively
large uncertainties on these bins.

Figure~\ref{f:fcomp} compares the values of $f$ obtained from the
three different methods.  In both panels, the horizontal axis is the
best-fit unweighted $f$ and the vertical axes are the best-fit
weighted absorbed (left panel) and weighted, unabsorbed (right
panel). All points correspond to a single model viewed at $60^\circ$
from the surface normal.  Blue circles show models with spectral peak
in $I_\nu$ above 0.5 keV and red symbols fall below.  For models
peaking above 0.5 keV the best-fit $f$ values are almost always in
good agreement.  Below 0.5 keV, models tend to be more strongly
affected by the presence of edges, particularly for lower values of
$f$.  The weighted unabsorbed model finds slightly larger values of
$f$ (up to $f \sim 2$) due to the effect described above.  The
absorbed model is even more strongly affected with harder spectra
tending to provide $f$ values higher than the unweighted fit and
softer spectra giving lower values of $f$.  In summary, $f$ can be
estimated rather robustly for spectra where the color-corrected
blackbody provides a good fit.  In contrast, $f$ is sensitive to the
fitting method when the color-corrected blackbody is a poor fit.

\subsection{Spectral Hardening Variations}

We adopt the results of the unweighted fitting method to consider the
evolution of $f$ with $T_{\rm eff}$, $Q$, and $m_0$.  As discussed in
section~\ref{fitting}, the best-fit $f$ is relatively insensitive to
the fitting method, when $I_\nu$ peaks above 0.5 keV.  Models that
peak below 0.5 keV only occur for lower $T_{\rm eff}$ and moderate to
high values of $Q$.  In disks with accretion rates above about a few
percent of the Eddington rate, these annuli correspond to relatively
large radius and have relatively little impact on the global disk
spectrum.  The global disk spectrum is primarily determined by the
spectral shape of the hottest, inner disk annuli, where a
color-corrected blackbody is a more suitable approximation and $f$
is relatively independent of fitting method.

Figure~\ref{f:table3} shows the variation of $f$ as a function of
$T_{\rm eff}$ and $Q$ for three different values of $m_0$.  The best
fit $f$ is shown for an inclination near $60^\circ$, with $f$ values
denoted by color.  To show the trend we fix the maximum of the
color bar at $f=4$, but even higher values occur. Except
for relatively low values $f \lesssim 1.4$, where the color-corrected
blackbody tends to be a poor fit, contours of constant $f$ tend
to run diagonally in the $Q$ -- $T_{\rm eff}$ plane, with $f$ generally
increasing with higher $T_{\rm eff}$ or lower $Q$.  Comparison of different
panel also shows a dependence on $m_0$, with lower $m_0$ corresponding
to harder spectra at the same $T_{\rm eff}$ and $Q$.  

The black dashed lines show the values of $Q$ and $T_{\rm eff}$
computed at the radius where $T_{\rm eff}$ reaches its maximum value in a relativistic accretion disk
model for various values
of the accretion rate and mass for a non-spinning black hole \citep{ShakuraSunyaev1973,NovikovThorne1973}. For black hole
spin $a=0$, $R_{\rm max} = 9.5 r_g$, where $r_g = G M/c^2$ is the gravitational radius and $G$ is Newton's constant.  We can then compute $T_{\rm eff}$, $Q$, and $m_0$ in the radiation pressure dominated limit via
\begin{eqnarray}
T_{\rm eff} & = &\left(\frac{1.5 c^3 \dot{m}}{\sigma_{\rm sb}\kappa_{\rm es} \eta r_g r_{\rm max}^3}\right)^{1/4} T_{\rm rel}(a,r_{\rm max}),\label{eq:teff}\\
Q & = & \frac{c^2}{r_g^2 r_{\rm max}^3} Q_{\rm rel}(a,r_{\rm max}),\label{eq:q}\\
m_0 & = & \frac{32 \eta r_{\rm max}^{3/2}}{27\kappa_{\rm es} \alpha \dot{m}} m_{0,\rm rel}(a,r_{\rm max}).\label{eq:m0}
\end{eqnarray}
Here $\dot{m}$ is the accretion rate scaled to $\dot{M}_{\rm Edd} = 4 \pi c r_g/(\kappa_{\rm es} \eta)$, $\kappa_{\rm es}$ is the electron scattering opacity, $\sigma_{\rm sb}$ is the Stefan-Boltzmann constant, $\eta$ is the spin dependent radiative efficiency, $\alpha$ is the stress prescription parameter and $r_{\rm max}=R_{\rm max}/r_g$. $T_{\rm rel}$, $Q_{\rm rel}$, and $m_{0,\rm rel}$ are relativistic correction factors \citep{NovikovThorne1973}. The diagonal lines in each panel correspond (from left to right) to $\dot{m} = 0.01$, 0.1, and 1.  The horizontal lines correspond
(from top to bottom) to $M/M_\odot = 10,\ 10^3,\ 10^5,\ 10^7$.  

Since the spectral hardening factor of a global disk model correlates closely
with the value of $f$ in the hottest annulus, this provides a
good estimate of how $f$ will vary for a global disk.  Increasing
the black hole spin shifts the diagonal lines to the right (i.e.
towards higher $T_{\rm eff}$ for a given $M$ and $\dot{m}$.
In the left panel, we also show the variation of $T_{\rm eff}$
and $Q$ with radius in an accretion disk for $M=10 M_\odot$ and
$\dot{m}=0.1$ as a white, dashed curve.  This shows that $f$ is generally highest
in the annuli with the highest $T_{\rm eff}$.

Figure~\ref{f:table2} shows the same quantities, but for lower values
of $m_0$.  Here the ranges of $T_{\rm eff}$ and $Q$ with valid
spectral models are narrower due to difficulties with convergence at
these these lower surface densities.  The diagonal and horizontal
dashed lines have the same meaning as in Figure~\ref{f:table3}, but
only cover masses of $M/M_\odot = 10$ and 1000.  Comparison with
Figure~\ref{f:table3} shows an accelerating trend for $f$ to increase
with decreasing $m_0$.  Equation (\ref{eq:m0}) shows that the value of $m_0$ in the $\alpha$-disk model
is a function radius, mass, spin, and accretion rate.  It is also
sensitive to the details of the assumed stress
prescription \citep{DoneDavis2008} and thus likely to be
the least robustly estimated parameter.

Inspection of the contours of constant $f$ in Figure~\ref{f:table3}
shows that they tend to follow the lines of constant $\dot{m}$,
although not precisely.  As we move along one of these lines from the
upper right to the lower left, mass increases and the spectral
hardening factor increases.  Hence, we expect supermassive black holes
to be associated with larger spectral hardening factors at the same
Eddington rate, consistent with earlier inferences \citep{Doneetal2012}.

Figures~\ref{f:table3} and \ref{f:table2} show that $f$ remains fairly modest ($f \lesssim 2$) except for models on the extreme edge of the distribution corresponding to the lowest $Q$ for a given $T_{\rm eff}$.  The rapid increase in $f$ occurs roughly at constant Eddington ratio, but not precisely.  Spectra on the high mass end (lower $T_{\rm eff}$ and lower $Q$) tend to have larger $f$ than those at the same Eddington ratio at lower masses.  Also, the transition to large $f$ depends on $m_0$, with lower surface densities having harder spectra.  In fact, we only have a few models with $f \gtrsim 2$ for models with $m_0 \ge 10^5 \; \rm g/cm^2$ for the parameters explored here.

The black solid curves in Figures~\ref{f:table3} and \ref{f:table2} approximately demarcates the transition to extreme spectral hardening.  This curve corresponds to the relation
\begin{equation}
\log Q = \log Q_0 + 7.5 \log T_{\rm eff} - 2.125 \log m_0.
\end{equation}
Here
\begin{equation}
Q_0 = \frac{m_p^2 \kappa_{\rm es}^{7/8} \sigma_{\rm sb}^2}{\eta_0 c},
\end{equation}
where $m_p$ is the proton mass.  We also assume that the frequency integrated free-free emissivity corresponds to $\eta_{\rm ff} \approx \eta_0 T^{1/2} \rho^2 m_p^{-2}$, where $\rho$ and $T$ are suitably averaged densities and temperatures.  This relation can be derived by setting
\begin{equation}
\eta_{\rm ff} H = \sigma_{\rm sb} T_{\rm eff}^4.\label{eq:starved}
\end{equation}
Here $H$ is the disk scale height, which evaluates to
\begin{equation}
H = \frac{\kappa_{\rm es}\sigma_{\rm sb} T_{\rm eff}^4}{c Q},
\end{equation}
in the radiation pressure dominated limit of an $\alpha$-disk model \citep{ShakuraSunyaev1973}. We approximate $\rho \simeq m_0/H$ and $T \simeq (\kappa_{\rm es} m_0)^{1/4} T_{\rm eff}$ to evaluate $\eta_{\rm ff}$.

Equation~(\ref{eq:starved}) is an approximate condition that accretion disk produce enough photons per unit volume to provide the flux that is required due to losses of gravitational energy and work done by stresses in the disk.  When the integrated emissivity is lower than the radiative flux, the disk must get hotter than the standard $T \sim \tau^{1/4} T_{\rm eff}$ profile to increase the emissivity.  In this regime the disk models tend toward near isothermality with $T \gg T_{\rm eff}$.  This ``photon starved'' limit occurs for $Q$ and $T_{\rm eff}$ values falling to the lower left of the black solid curve.  In this regime, $f \sim T/T_{\rm eff} \gg 1$, leading to large spectral hardening.  Since the emissivity is strongly dependent on density, this transition happens at lower $T_{\rm eff}$ for a given $Q$ as we go to smaller values of $m_0$.  In other words, disks with low surface density become photon starved most easily.

We emphasize that the large range of $Q$, $T_{\rm eff}$, and $m_0$ allows us to rule out other reasons for the transition to rapid spectral hardening, which would give a different transition curve.  Conditions based on the effective optical depth $\tau_{\rm eff} \simeq \sqrt{3 \kappa_{\rm es}\kappa_{\rm abs}} m_0$ or simply the ratio $\kappa_{\rm es}/\kappa_{\rm abs}$ give different scalings in the $T_{\rm eff} - Q$ plane that do not explain the spectral hardening we infer.

\begin{figure}[h]
\includegraphics[width=9cm]{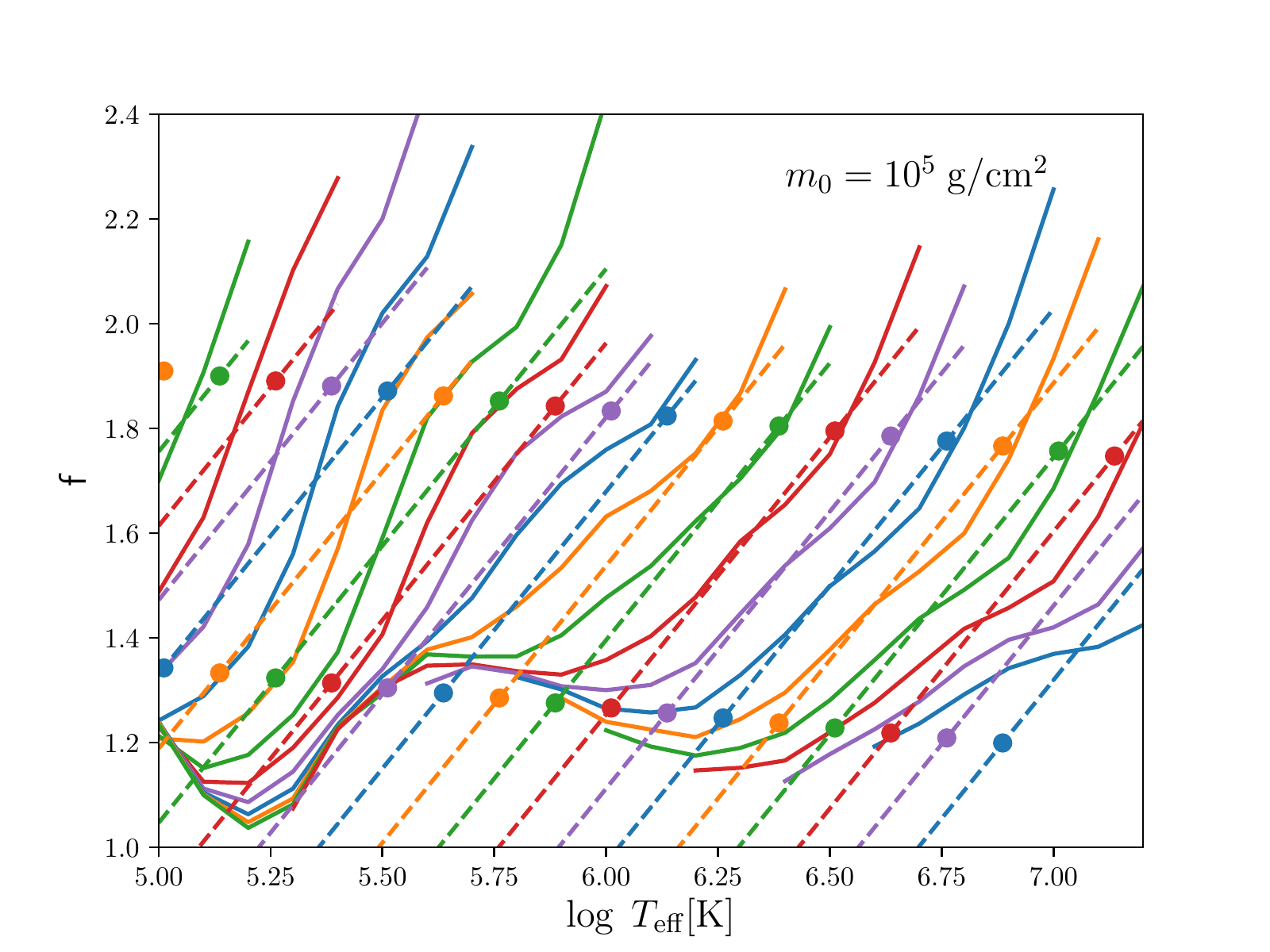}
\caption{Spectral hardening factor for $m_0 =10^5 \; \rm g/cm^2$.  Each set of 19 curves represents a different value of $Q$, running from $Q=10^{-8} \; \rm s^{-2}$ to $10^{10} \; \rm s^{-2}$ incremented in powers of 10. Solid curves show the values of $f$ derived from our unweighted fitting method while dashed curves represent the linear approximation in equation (\ref{eq:fitmod}).  Curves are color-coded so that nearly adjacent curves with the same color represent the same $Q$ value.  The circles show where $\dot{m}=0.01$ and 1 for the linear relation. \label{f:fit}.}
\end{figure}

\subsection{Approximate Spectral Hardening Variation}

In general, the spectral hardening factor displays a complicated variation with $T_{\rm eff}$, $Q$, and $m_0$ that is is not easily reproduced by a simple fitting function.  However, if we restrict our attention to regions of parameter space where bound-free edges have relatively modest impact on the spectrum and photon starvation is not a factor, the variation of $f$ is simpler.  We can approximately accomplish this for a non-spinning black hole by restricting our attention regions of parameter space where $m_0 \ge 10^3 \; \rm g/cm^2$ and $0.01 < \dot{m}  < 1$.  Then we find that $f$ can be crudely approximated by linear fit to the log of $T_{\rm eff}$, $Q$, and $m_0$, with best-fit parameters
\begin{eqnarray}
f & = & 1.74 + 1.06 (\log T_{\rm eff}-7) - 0.14 (\log Q - 7)\nonumber \\
& & -0.07 (\log m_0 - 5),\label{eq:fit}
\end{eqnarray}
where all quantities are evaluated in cgs units.

Figure~\ref{f:fit} shows a comparison of this linear relation and our best fit spectral hardening factors for $m_0 =10^5 \; \rm g/cm^2$.  Each set of 19 curves represents a different value of $Q$, running from (left-to-right) $Q=10^{-8} \; \rm s^{-2}$ to $10^{10} \; \rm s^{-2}$ incremented in powers of 10.  Each set of two curves is a color-coded pair, with solid curves representing the best fit $f$ and  dashed lines corresponding to equation (\ref{eq:fit}).  We see that the approximation has problems when $\dot{m} \sim 0.01$, where bound-free edges begin to become important.  We also see that the linear relation works best for intermediate values of $Q$, not rising steeply enough with $T_{\rm eff}$ for low $Q$ (higher black hole mass) and rising too steeply for high $Q$ (lower black hole mass).

One can use equations (\ref{eq:teff})-(\ref{eq:m0}) to replace $T_{\rm eff}$, $Q$, $m_0$ with $\dot{m}$, $M$, $\alpha$ in equation (\ref{eq:fit}). Assuming a non-spinning black hole and that radiation pressure always dominates, we find
\begin{eqnarray}
f & \simeq & 1.48+0.33 (\log \dot{m} + 1) +0.02 (\log [M/M_\odot] - 1) \nonumber\\
 & & +0.07(\log \alpha + 1).\label{eq:fitmod}
\end{eqnarray}
We see that $f$ increases with $\dot{m}$, $M$, and $\alpha$, depending most sensitively on $\dot{m}$ and least sensitively on $M$.

We caution that (\ref{eq:fitmod}) is presented only to provide a sense of the spectral hardening evolution.  There are uncertainties in the underlying model, an assumption that radiation pressure dominates, questions about the robustness of our fitting procedure to estimate $f$, and the assumption that $f$ for the full disk is well-approximated by the annulus with the highest $T_{\rm eff}$. If we accept these caveats, figure~\ref{f:fit} still shows the linear relation only approximately holds over a limited range in accretion rate, mass and disk surface density.  It also only applies for $a=0$.  Recomputing equations (\ref{eq:teff})-(\ref{eq:m0}) for other $a$ values will yield the same scalings with $\dot{m}$, $m$, and $\alpha$, but with a higher spectral hardening factor.  For example, if $a=0.9$, 1.48 is replaced 1.56 in equation (\ref{eq:fitmod}).

\section{Discussion and Conclusions}

\subsection{Origin of extreme spectral hardening}

Our results suggest that extreme spectral hardening can occur in black hole accretion disks, but it generally only occurs in regions of the disk with high effective temperature if the disk surface density is sufficiently high. For standard disk models, this requires super-Eddington accretion rates, which are somewhat rarely inferred in X-ray binaries and rather infrequently and unreliably inferred in AGN.  The exception is when the disks surface density is relatively low.  If the mass per unit area in the disk is below about 1000 $\rm g/cm^2$, we find that high spectral hardening factors occur at more modest effective temperatures.

In a $\alpha$-disk model, surface density tends to scale inversely with $\alpha$ and accretion rate. For $a=0$, equations (\ref{eq:m0}) evaluates to $m_0=2 \times 10^5 (0.1/\dot{m})(0.1/\alpha) \; \rm g/cm^2$.  Surface densities of $\sim 1000 \; \rm g/cm^2$ only occur for $\alpha \sim 1$ and $\dot{m} \sim 1$ for $a=0$. However, spinning black holes can have surface densities that are lower for the same $\dot{m}$ and $\alpha$. Since high Eddington ratios seem to be rare, we expect most observed accretion disks to be consistent with more moderate $f \lesssim 2$.  This roughly justifies the standard assumptions of $f \sim 1.7 - 1.8$ in many analyses.  The lack of strong evidence for extreme spectral hardening in most high/soft state X-ray binaries \citep[except when moving into or out of the high state]{Dunnetal2011} also suggests surface densities consistent with lower values of $\alpha$ or some sort of alternative stress prescription \citep{DoneDavis2008}.

Some observations do infer large spectral hardening factors in the hard state of X-ray binaries. \cite{Dunnetal2011} find evidence for larger $f$ as the disk transitions in or out of the high/soft state.  \citet{Salvesenetal2013} study the low hard state in numerous observations of GX339-4 and conclude the disk requires $f \sim 3$ in this state.  Large values of $f$ have been attributed to energy being deposited in the corona rather than the disk based on results by MFR00, but we believe this is a misinterpretation of those results, as we discuss below.  If the spectral hardening factor in these disks are as high as has been inferred, our results suggest this mostly likely corresponds to a reduction in the disk surface density.  We emphasize that the disk need not be optically thin for such a transition to occur.  The photon starvation limit described here can easily happen in disks with Thomson optical depths of more than 100.  Hence, reflection signatures (e.g Fe lines) are still expected to be present if the disk is irradiated by a corona.  It must be emphasized that our results assume a single temperature for electrons and protons.  Two-temperature accretion disk models (e.g. advection dominated flows or ADAFs) may be more relevant as we approach lower surface densities \citep{NarayanYi1994,Narayan1996}, which require large inflow velocities.  It is not clear whether such only moderately optically thick single temperature flows arise naturally in a quasi-steady state accretion flow before transitioning to optically thin, two-temperature flows.

\subsection{Comparison with previous work}

The primary motivation for using a color-corrected blackbody comes
from ST95 who showed both that the color-corrected blackbody is a
suitable approximation to the Comptonized disk emission and concluded
that $f$ varies only over a rather narrow range in accreting black
hole sources. This conclusion is consistent with the relatively narrow range of parameter space considered by ST95.  Our values of $f$ qualitatively agree with theirs in the same range. Since our results are based on the same models as
D05, they trivially agree with D05 over the same range.
The key difference here is that we extend the exploration of the spectral
hardening factor to a much wider range of effective temperatures and
characteristic densities than considered in D05.  Hence we find a broader range of hardening factors
but only because we consider a much broader parameter space.

At lower Eddington ratios, the results of ST95 were contradicted by
the work of MFR00. MFR00 adopted
the the model of \citet{SvenssonZdziarski1994}, which uses a two-zone model to approximate the
impact of a fraction of the energy being dissipated in a corona rather
than in the underlying disk.  In
the MFR00 calculations the emission arising from the
corona is not included.  In particular, there is no corresponding
irradiation of the underlying disk when a large fraction of the
dissipation is assumed to occur in the corona.  Hence, the primary
impact is that the flux $F$ in the model is multiplied by $(1-\chi)$,
where $\chi$ is the fraction of energy assumed to be lost to the
corona.\footnote{Since we have already reserved $f$ for the spectral hardening factor, we use $\chi$ in place of $f$ used by MFR00 for the fraction of energy lost to a corona.}

\begin{figure}[h]
\centering
\includegraphics[width=4.2cm]{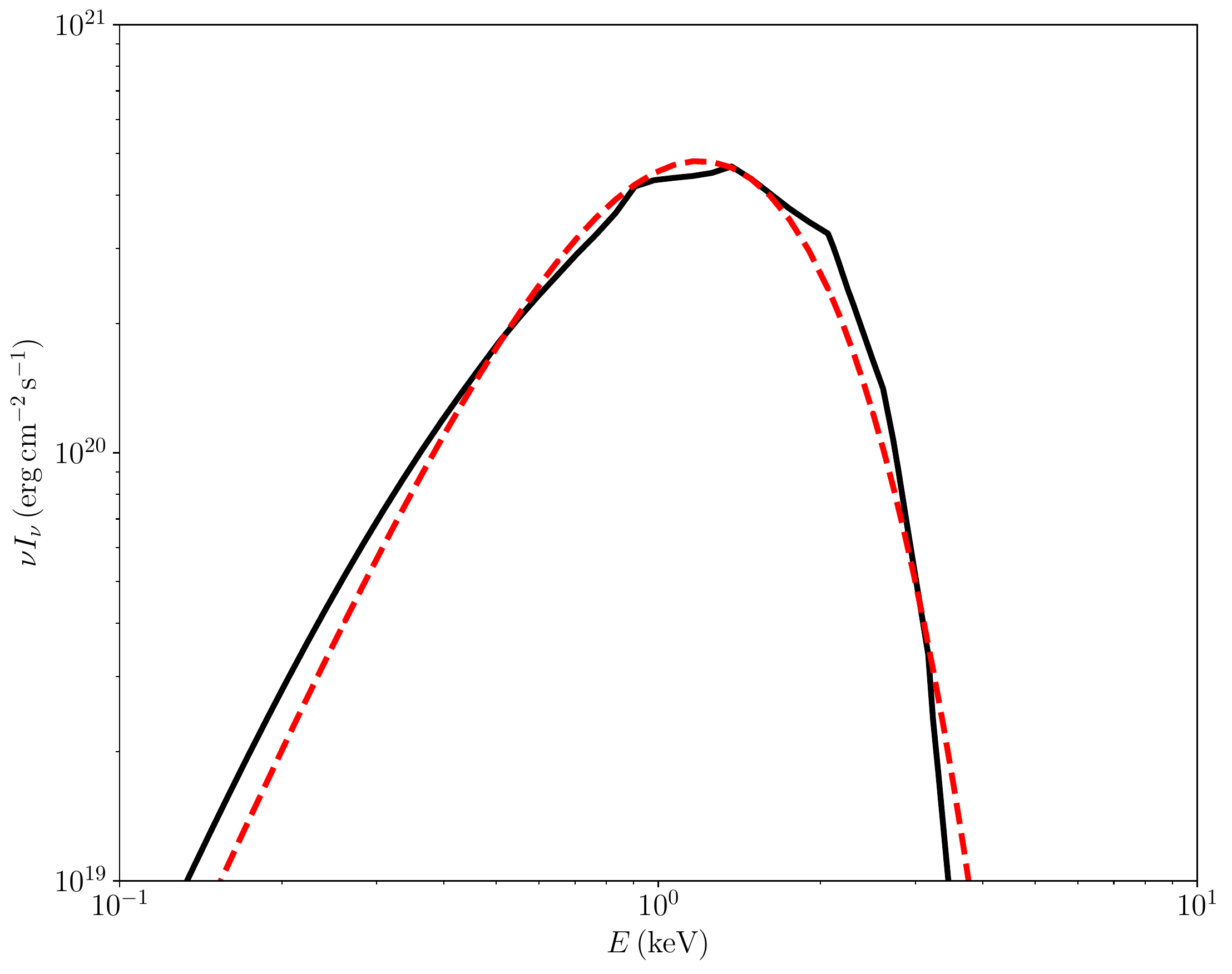}
\includegraphics[width=4.2cm]{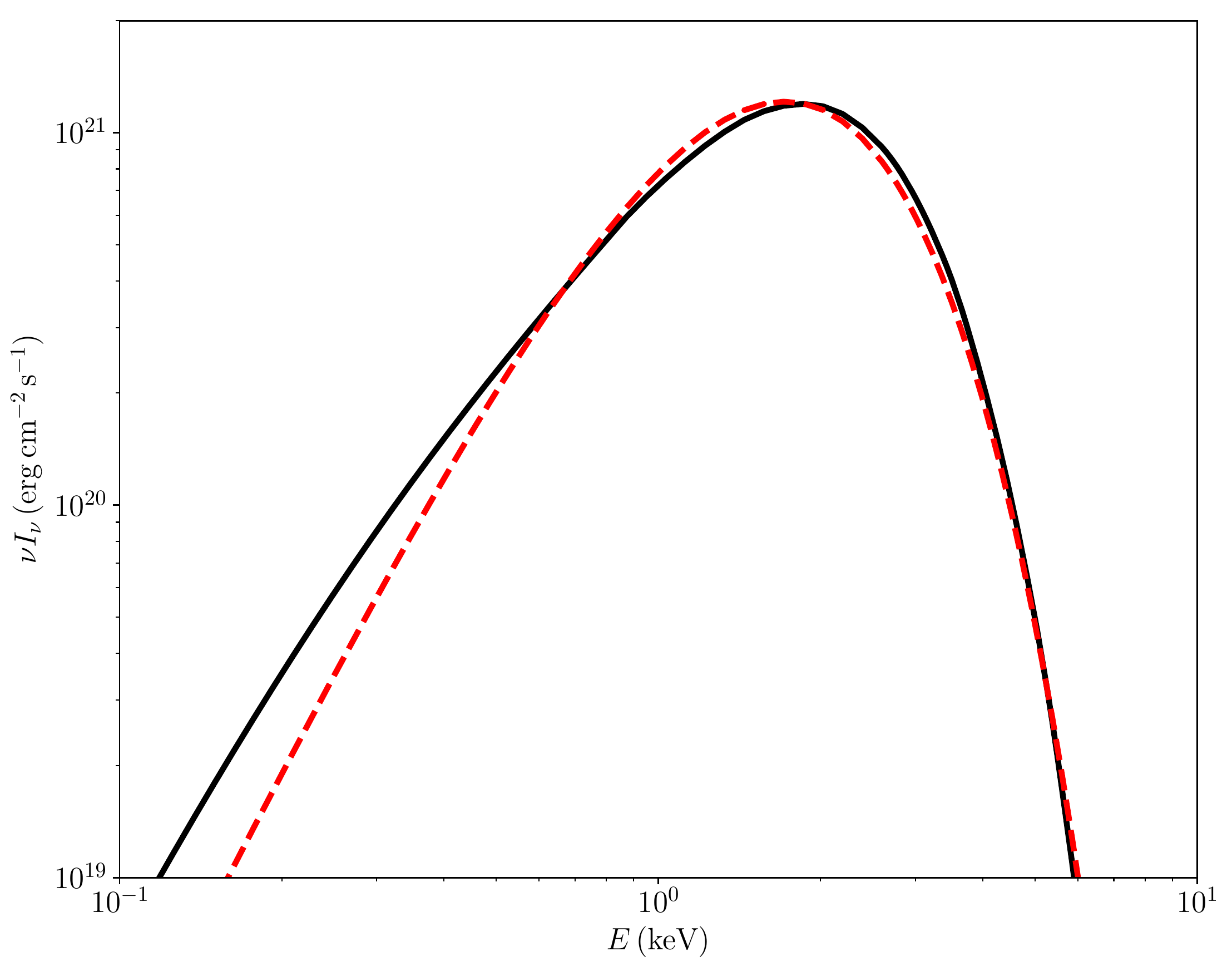}

\caption{Spectrum (solid, black) and best-fit color-corrected blackbody (dashed, red) viewed at an inclination of $60^\circ$.  Model parameters in the left panel correspond to $T_{\rm eff}=2.5 \times 10^6 \; \rm K$, $Q= 10^5 \; \rm s^{-2}$, and $m_0= 10^5\; \rm g/cm^2$ with best-fit $f=1.43$.
The right panel shows $T_{\rm eff}=3.2 \times 10^6 \; \rm K$, $Q= 10^5 \; \rm s^{-2}$, and $m_0= 10^4\; \rm g/cm^2$ with best-fit $f=1.62$.}\label{f:compmer}
\end{figure}

MFR00 find their highest spectral hardening for models with large $\chi$, corresponding to low values of $F(1-\chi)$, or low $T_{\rm eff}$.  Although they emphasize that large $f$ occurs for $\chi \gtrsim 0.5$,
MFR00 do not present enough spectral calculations to discriminate between the effect of increasing $\chi$ or simply lowering the flux $F$. For the two pairs of models with the same $F(1-\chi)$ but differing values of $F$ and $\chi$ (their S3/S4, and S8/S9) the resulting spectral hardening factors seem to depend primarily on the product $F(1-\chi)$, rather than $\chi$ or $F$ independently.  Hence, we assume that large $f$ in their models is not an effect specifically of transferring energy to a corona (i.e. having $\chi \ne 0$) but a general consequence of a lower $T_{\rm eff}=(F(1-\chi)/\sigma_{\rm sb})^{1/4}$ in models with larger $\chi$.  In contrast, we almost never find large spectral hardening for such low effective temperatures and our models are clearly in contradiction with those of MFR00, as noted previously in D05.

Due to the significant differences in the calculations, it is difficult to unambiguously determine the source of the discrepancy. D05 speculate that it may have something to do with the assumption of constant density in the MFR00 calculations or effect of bound-free opacities that are present in the TLUSTY calculations but not in those of MFR00. The S11 model of MFR00 has the highest spectral hardening, with $f=2.68$ and $\chi=0.8$  Using their expressions for $\rho_0$, $h$, and $F_0$ (their equations [2]-[6]), we estimate $m_0=h \rho_0 R_{\rm S}=2.3 \times 10^5 \; \rm g/cm^2$, $Q=2.4 \times 10^5 \; \rm s^{-2}$, and $T_{\rm eff}=2.3 \times 10^6 \; \rm K$ at $r=6 R_{\rm S}$ where $R_{\rm S}$ is the Schwarzschild radius.  We find $f=1.43$ for our model with similar parameters, which is shown in the left panel of Figure~\ref{f:compmer}.  For comparison, we also compute a model with the same parameters as S11, but setting $\chi=0$ so no energy is dissipated in a corona.  In this case, we find $m_0=9.1 \times 10^3 \; \rm g/cm^2$, $Q=2.4 \times 10^5 \; \rm s^{-2}$, and $T_{\rm eff}=3.5 \times 10^6 \; \rm K$ at $r=6 R_{\rm S}$.  Our closest model is shown in the right panel of Figure~\ref{f:compmer} and has a best fit spectral hardening factor is $f=1.62$. Hence, for this simplified prescription where the irradiation by the corona is not included, a lower dissipation of energy in the disk actually makes the disk spectrum {\it softer}, in conflict with the conclusions of MFR00.

Figure~\ref{f:compmer} shows our $\chi=0.8$ model has moderately strong features due to bound-free opacities.  So, it is plausible that the presence of bound-free opacity in our models is keeping the disk closer to a blackbody spectrum than in the MFR00 calculations, which neglect bound-free opacity.  In contrast, our $\chi=0$ model is hotter and has a lower surface density, leading it to be in regime where bound-free opacity has little effect and Compton scattering dominates.  Another concern is the treatment of Compton scattering in these calculations.  They define coherence radius beyond which electron scattering is treated as coherent and the effects of Comptonization (which tend to soften the spectra) are ignored.  For the highest $\chi$ models, the annuli computed without Compton effects include the hottest annuli in these disks.  The coherence radius was chosen based on a Compton $y$-parameter condition.  If this choice was insufficiently conservative, it is conceivable that neglecting Compton scattering in these annuli could have led a modified blackbody spectrum that is harder than if Compton scattering had been included in all calculations, but this is only speculation on our part.

%Finally, we note that a somewhat narrow range of spectral hardening factors are also found in the study of bursting neutron star atmospheres.  Although, the gravity and dissipation profiles are different, the same Compton scattering effects dominate both systems.  We have performed a small number of test calculations with constant gravity and flux, finding that TLUSTY provides approximate agreement with \cite{Suleimanovetal.2011} who use a similar treatment for Compton scattering.  Most notable here is that extreme spectral hardening ($f > 2$) is not seen in neutron star systems, consistent with the large effective surface density in neutron star atmospheres.

\subsection{Soft X-ray Excess}

Many AGN show a soft X-ray excess, where the emission below $\sim 1$ keV exceeds the extrapolation of the hard $>2$ keV power law continuum.  Proposals for the origin of this soft X-ray excess include relativistically smeared reflection \citep{Crummyetal2006} or absorption \citep{GierlinskiDone2004}.  A third possibility is that soft excess is continuum emission from the accretion disk \citep[e.g.][]{Doneetal2012}.  In order for the emission scenario to work with standard disk models one requires lower mass black holes ($M \lesssim {\rm few} \times 10^6 M_\odot$) and Eddington ratios near unity to obtain emission at energies as high as $\sim 1$ keV.  In higher mass objects, it seems that the inner disk will be too cool to explain soft-excesses, suggesting that some warm Comptonizing region exists with characteristic temperatures of $\sim 1$ keV \citep{Czernyetal2003}.

The results here suggest that a low surface density disk might also contribute to the soft X-ray emission in AGN.  Given the theoretical uncertainties underlying radiation dominated regions of accretion flows, it seems plausible that real accretion flows might have surface densities that are optically thick to electron scattering but still well into the photon starved limit discussed here.  The primary question is whether such a configuration would naturally give turnovers in the vicinity of $\sim 1$ keV and be consistent with the variability properties of the soft X-ray emission.  

\subsection{Model Uncertainties}

The primary caveats to our conclusions are that we have to make a number of underlying assumptions in determining the disk vertical structure and radiation transfer.  Key uncertainties include the vertical distribution of dissipation, effects of magnetic pressure support, and inhomogeneities \citep{bla06,dav09,TauBlaes2013}.

We also neglect the effects of bound-bound transitions and irradiation, due either to a corona or returning radiation from the opposite side of the accretion disk, which seems substantial in ray tracing calculations of black hole accretion simulations \citep{Schnittmanetal2016}.  Models of the accretion disk that including bound-bound transitions and irradiation \citep{Rozanskaetal2011} do produce notable emission lines but do not seem to yield substantial differences in the underlying continuum.
 
\subsection{Conclusions}

We utilize a large table of accretion disk annuli spectra generated with the TLUSTY code to study the variation in spectral hardening over a wide range of accretion disk parameters.  We perform a series of fits with color-corrected blackbody models to compute the spectral hardening factor $f$.  Consistent with most previous work, we find the $f$ varies over a somewhat narrow range $f \sim 1.4-2$ for the parameters in the innermost regions of black hole accretion disks for typical X-ray binary accretion rates ($0.03 \lesssim \dot{m} < 1$) and masses. Consistent with previous results, we find $f$ depends most sensitively on the accretion rate, with higher $f$ for higher $\dot{m}$.

Our results also show that extreme spectral hardening (defined here as $f > 2$) can be found in accretion disks which become photon starved.  This usually only occurs for relatively high effective temperature unless the disks have lower surface densities than commonly inferred with $\alpha$-disk models.  We suggest that observational evidence for higher values of $f$ are therefore best explained as coming from accretion disk with lower surface densities, although such disks can still be quite Thomson thick.  We argue that claims that high values of $f$ can be attributed to dissipation of a substantial fraction of the disks energy in a corona are likely based on an incorrect interpretation of previous work that is contradicted by our results.  We believe that our calculations are more credible due to our inclusion of bound-free opacity sources, more careful treatment of disk vertical structure, and Compton scattering effects.

Finally, our results show that spectral hardening tends to be larger for higher mass black holes when evaluated at the same Eddington ratio. For lower supermassive black hole masses $M \sim 10^6 M_\odot$ accreting near the Eddington limit, the inner accretion disk in $\alpha$-disk model becomes hot enough to produce substantial soft X-ray radiation.  If the surface densities are smaller than predicted by the standard $\alpha$-disk model or $\alpha \sim 1$, it is possible that the bulk of the soft X-rays excess emission comes from such photon starved regions of disks.

\acknowledgments

We thank Omer Blaes, Chris Done, Ari Laor,  Ramesh Narayan, and Greg Salvesen for useful conversations.  This work relies heavily on models originally computed in collaboration with Omer Blaes and Ivan Hubeny. S.W.D. acknowledges support from NASA Astrophysics Theory Program grant 80NSSC18K1018 and an Alfred P. Sloan Research Fellowship.

\bibliography{ms}

\end{document}